\documentclass[]{ceurart}
\usepackage{listings}
\usepackage{graphicx}
\usepackage{subcaption}
\begin{document}

\copyrightyear{2026}
\copyrightclause{Copyright for this paper by its authors.
  Use permitted under Creative Commons License Attribution 4.0
  International (CC BY 4.0).}

\conference{Joint Proceedings of the ACM Intelligent User Interfaces (IUI) Workshops 2026, July 13-16, 2026, Paphos, Cyprus}

\title{Talking to Your Data: Exploring Embodied Conversation as an Interface for Personal Health Reflection}

\author[1]{Nikola Kovacevic}[%
orcid=0000-0002-1337-7670,
]

\author[1]{Bastien Husler}[%
orcid=0009-0009-3655-1916,
]

\author[1]{Di Zhuang}[%
orcid=0009-0009-2693-9765,
]

\author[1]{Rafael Wampfler}[%
orcid=0000-0003-0158-1305,
email=rafael.wampfler@inf.ethz.ch
]

\author[1]{Barbara Solenthaler}[%
orcid=0000-0001-7494-8660,
email=solenthaler@inf.ethz.ch
]
\cormark[1]

\address[1]{Department of Computer Science, ETH Zurich,
  Zurich, Switzerland}

\cortext[1]{Corresponding author.}

\begin{abstract}
  Personal health data from wearables are typically presented through dashboards of charts and summary statistics, requiring users to actively interpret patterns and implications. We explore an alternative interaction paradigm: engaging with personal health data through an embodied conversational agent that facilitates objective data reflection in dialogue with the user.
  We present a system that combines lightweight preprocessing of wearable data with a Unity-based embodied character. Internally, the system follows a dual-agent design in which an Observer agent extracts descriptive statistics and temporal trends, and a Presenter agent communicates these findings through "spoken statistics," intentionally refraining from clinical advice to isolate the impact of the interaction modality.
  We evaluate this approach through a simulated-self user study (N=5) using a within-subject design. Participants adopted health personas and goals derived from the LifeSnaps dataset to compare traditional dashboard exploration with embodied conversational reflection. Our evaluation focuses on perceived understanding, the specificity of generated actions, and the cognitive shift from passive viewing to active sensemaking. The paper contributes a functional prototype, a design pattern for objective health data narrative generation, and early empirical insights into how embodiment affects the interpretation of personal health metrics.
\end{abstract}


\maketitle

\section{Introduction and Background}
\label{sec:introduction}

Wearable devices and mobile sensors routinely track heart rate, physical activity, and sleep, making personal health data widely available.
Despite this, long-term engagement with health and wellness applications remains low: users often disengage after weeks or months, even when the data could support reflection, learning, or lifestyle adjustments~\cite{Demiris2016,Demiris2022,Lai2017}.
A central reason lies in how health data is presented. Most systems rely on dashboards of charts and numerical summaries.
While effective for expert analysis, these representations place a high cognitive burden on users, require statistical literacy, and offer limited support for interpretation or reflection.
Reviews of patient-facing health visualizations consistently report challenges related to comprehension, relevance, and sustained engagement~\cite{Turchioe2019,Chan2024}.
Prior work in personal informatics further shows that breakdowns in engagement often stem not from data quality, but from difficulties in making data meaningful in everyday life~\cite{Li2010,Consolvo2008,Epstein2015}.

Recent work has explored conversational systems as an alternative interface for engaging with health data.
Large language models have been used to interpret wearable signals, generate personalized recommendations, and support conversational exploration of personal health information~\cite{Kim2024,Cosentino2024,Wang2025,Fang2024}.
These systems demonstrate that dialogue can lower barriers to data access and help users articulate goals and questions.
However, most existing approaches remain primarily text-based and treat conversation as a delivery channel for insights rather than as an interactive experience in its own right.
As a result, interaction with health data often remains abstract and detached, even when mediated by conversational agents.

We argue that embodiment introduces a fundamentally different interaction modality for personal health data.
In this work, embodiment refers to a visually present conversational character that shares the user’s visual context and supports joint attention around the data, rather than to physically situated or full-body interaction.
Embodied conversational agents can convey social presence, attentiveness, and relational cues that are difficult to achieve with text or voice alone~\cite{Cassell2000,Paiva2017}.
Prior work in human--agent interaction and eHealth suggests that relational and affective qualities play an important role in user engagement and perceived support~\cite{Provoost2017,terStal2020}.
Yet, few health data systems integrate even lightweight data interpretation with embodied, multimodal interaction.
Consequently, health data is still experienced primarily as a report to inspect, rather than something users can engage with, reflect on, or discuss.
Conversational approaches to health reflection have also been explored through chatbots, voice assistants, and coaching systems for behavior change and chronic disease management (e.g., \cite{Bickmore2011,Provoost2017}); our work differs in examining how conversational narration operates in a shared-view setting alongside an existing dashboard rather than replacing it.

In this paper, we explore \emph{talking to your data} as an interaction paradigm for personal health information.
We present an embodied conversational system that allows users to discuss wearable-style health data with a virtual agent that explains trends, contextualizes values, and supports reflection through dialogue and simple visualizations.
The system follows a dual-agent architecture: an \emph{Observer} agent performs lightweight preprocessing to extract objective statistical insights from time-series data, while a \emph{Presenter} agent communicates these insights through embodied conversation.
The system is intentionally designed for consumer health reflection rather than clinical decision-making.

To investigate how embodied conversation changes users' experience of health data, we conduct a controlled, within-subject user study comparing two conditions: (1) a traditional dashboard-based interface, and (2) the same interface extended by an embodied conversational agent acting as a narrative interpreter of that same data, with both conditions operating on identical data and predefined health goals.
Our evaluation focuses on perceived understanding, actionability, motivation, and engagement, isolating the role of interaction modality.
Through this study, we aim to provide early empirical insights into how embodied, conversational interfaces may reshape the way people engage with personal health data.
As an exploratory pilot with a small, homogeneous sample (N=5), our findings should be interpreted as hypothesis-generating observations about interaction modality rather than generalizable evidence about user populations.



\section{System Design}
\label{sec:system_design}

Our goal is to explore how embodied, conversational interaction reshapes engagement with personal health data.
To this end, we designed a system that allows users to \emph{talk to their data} through an embodied conversational agent that explains and contextualizes wearable-style health information.
The system is intentionally non-clinical, focusing on reflective engagement rather than medical decision-making.
In this exploratory study, we examine how conversational narration and embodied presence influence users’ reflection on their data and the specificity of the actions they derive from it.

\subsection{Design Goals}
\label{subsec:design_goals}

The system was guided by three goals derived from prior work in personal informatics~\cite{Li2010} and embodied interaction~\cite{Cassell2000}: 

\begin{itemize}
    \item \textbf{G1: Support reflective engagement through embodiment.} Drawing on the "Computers are Social Actors" paradigm~\cite{Nass1994}, the agent provides social presence to frame data reflection as a collaborative dialogue rather than a solitary task.
    \item \textbf{G2: Lower the barrier to understanding personal health data.} To address known sensemaking challenges in health dashboards~\cite{Choe2014}, the system provides a conversational layer that explains patterns, reducing the cognitive burden of manual chart interpretation~\cite{Pirolli2005}. By verbally referencing trends and values, the agent supports sensemaking without requiring the user to navigate the visual information alone.
    \item \textbf{G3: Enable controlled comparison.} Following experimental design standards in IUI research, the system ensures that underlying insights remain constant across modalities to isolate the effects of the conversational interface.
\end{itemize}

\subsection{System Overview}
\label{subsec:system_overview}

\begin{figure}
    \centering
    \includegraphics[width=0.99\linewidth]{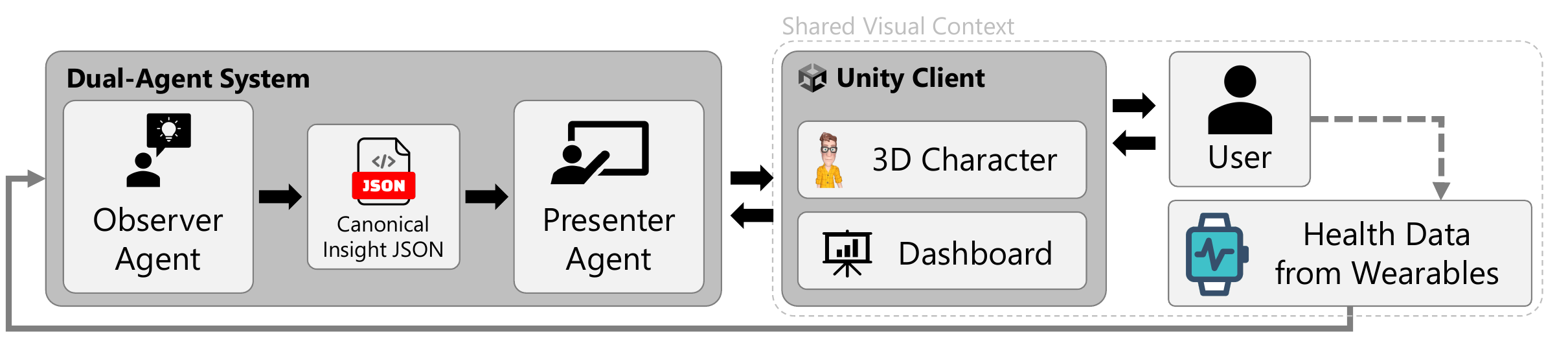}
    \caption{The Dual-Agent Architecture and Unity Integration. The system employs a decoupled design to ensure data-driven narrative reflection. The Observer Agent utilizes rule-based feature extraction to preprocess longitudinal health traces into a canonical \emph{Insight JSON} containing both raw metrics and natural-language insights. The Presenter Agent consumes this JSON to ground its responses, functioning as a "narrative bridge" that translates abstract data into a social dialogue. The Unity Client provides the shared visual context, featuring a static Dashboard for data visualization alongside a 3D Character that embodies the Presenter Agent's output to facilitate joint attention and co-interpretation.}
    \label{fig:system_overview}
\end{figure}

As depicted in Figure~\ref{fig:system_overview}, our system consists of two main parts, the dual-agent system and the unity client.
Operating on static wearable real-life data, the dual-agent system first preprocesses this data via the Observer agent, producing a canonical representation of statistics and insights.
It then passes the data to the Presenter agent which acts as a narration layer between the user and the data. 
The Unity-based client, on the other hand, serves as a conversational interface to the user.
It hosts an embodied character that synthesizes the presenter's messages to speech and animations, while a static dashboard displays the aggregated health data as charts, allowing the agent to refer to these visualizations in a shared visual context.

\subsection{Dual--Agent Architecture}
\label{subsec:dual_agent}
The motivation behind the dual-agent design is to separate data interpretation from user-facing presentation, which was shown to improve clarity and controllability in similar contexts~\cite{Fang2024,Wang2025}. We employ two separate agents, an Observer and a Presenter agent:

\paragraph{Observer Agent.} The Observer Agent performs a hybrid analysis of the raw data to extract structured insights.
First, it executes deterministic scripts to calculate descriptive statistics, correlations, and temporal trends using weighted linear regression with decay.
Second, it utilizes a Large Language Model (GPT-4o-mini) to identify non-obvious patterns, such as "rebound effects" after active days, based on the raw daily data and variance
\footnote{A detailed overview of the user study setup is available on \href{https://gitlab.inf.ethz.ch/prj-cgl/cgl-ai-character/talking-to-your-data.git}{our GitLab repository}.}.
The output is a structured \emph{Insight JSON} containing precomputed statistics and insights with associated confidence scores.

\paragraph{Presenter Agent.} The Presenter Agent transforms the Observer's insights into conversational explanations using OpenAI's GPT-4o-mini.
The agent is governed by a Strict-Grounding Prompting strategy~\cite{Lee2024}, constrained to only reference facts within the Insight JSON and prohibited from providing medical advice.
Responses are limited to a maximum of five sentences, emphasizing reflective questions and "spoken statistics" (e.g., using weekdays like "last Wednesday") rather than raw dates to maintain a natural conversational flow.

\subsection{User Interaction} 
\label{subsec:user_interaction}

Users interact with our system through a Unity client that provides two interfaces, a dashboard interface for displaying health data and an embodied conversational agent for speech-based interaction.

\paragraph{Dashboard Interface.} Through the dashboard, users explore their health data through static charts and labels similar to those displayed on wearable devices.
The dashboard is inspired by visuals from the Apple Watch and shows bar charts and an average per data type.
To facilitate a controlled interaction experiment as outlined in Section~\ref{sec:user_study}, the dashboard is additionally capable of showing information relevant to the user study such as a persona description and personal health goals.

\paragraph{Embodied Conversational Agent.} Users interact with a 3D upper-body avatar through speech by clicking on a microphone button.
The agent uses speech synthesis, with lip-synced animation and simple gaze behavior.
To maintain a focused comparison on the interaction modality, the agent does not directly manipulate the visualizations on the dashboard.
Instead, it engages in a "shared-view" dialogue, explicitly referencing the values and trends visible in the static charts (e.g., "As you can see on the sleep chart, last Thursday was quite low") to support the user's reflection.

\section{User Study} \label{sec:user_study}

To test our system and provide early insights into the \emph{talk to your data} paradigm, we conducted a controlled, mixed-design pilot study ($N=5$).
The study employed a "Simulated Self" protocol, where participants adopted a lifestyle persona to mitigate the privacy constraints of personal medical data and ensure comparability across participants while maintaining a realistic data-reflection task.

\begin{figure}[t] 
    \centering 
    \begin{subfigure}[b]{0.57\textwidth} 
        \centering \includegraphics[width=\textwidth]{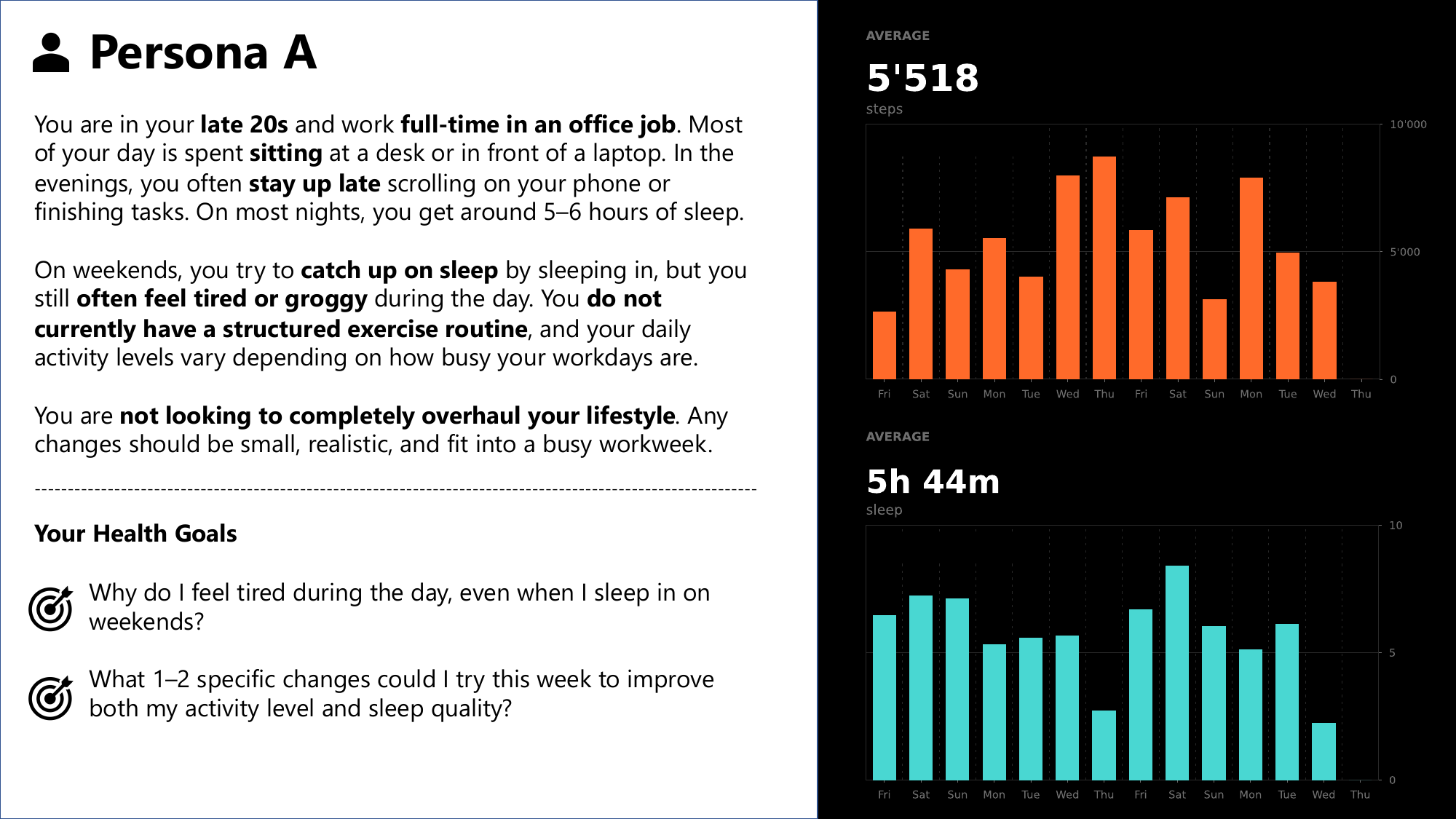} 
        \label{fig:persona_a} 
    \end{subfigure} 
    \hfill 
    \begin{subfigure}[b]{0.419\textwidth} 
        \centering \includegraphics[width=\textwidth, trim=4.5cm 0cm 4.5cm 0cm, clip]{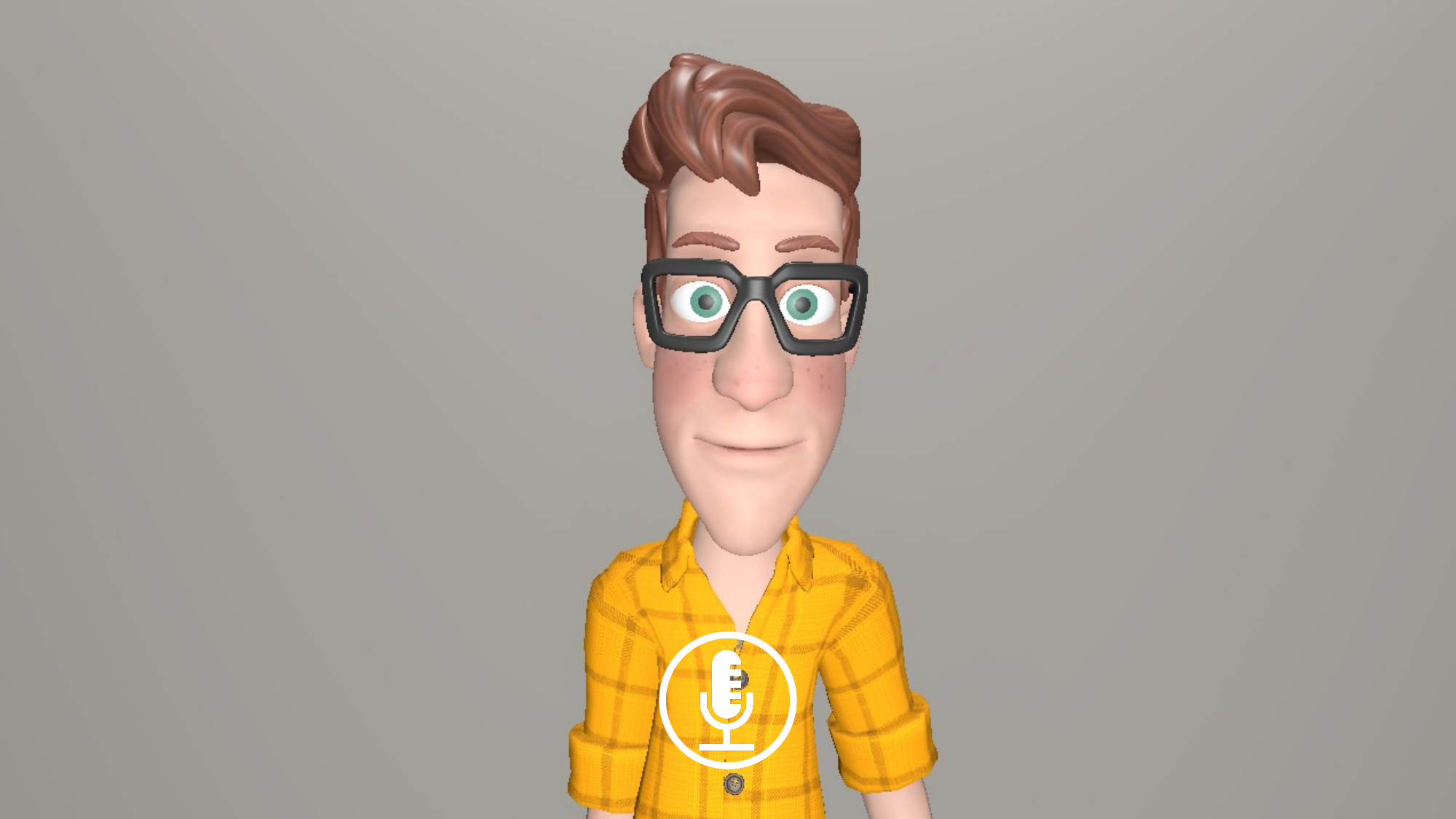}
        \label{fig:vincent} 
    \end{subfigure} 
    \caption{Screenshot of the User Interface. The left screen shows the health data charts alongside persona information and health goals specific to the user study. The right screen shows the embodied conversational agent during a conversation.} \label{fig:interface} 
\end{figure}

\subsection{Apparatus and Stimuli} \label{subsec:stimuli}

The study was conducted in a dedicated observation room using a three-screen configuration to minimize cognitive load during multi-modal interaction. Two large displays presented the health data and the system interface---either only the Dashboard (left screen) or both the Dashboard (left screen) and the Agent (right screen)---while a third display (a MacBook) was dedicated to psychometric questionnaires\footnote{\label{fn:repoRef}A detailed overview of the user study setup is available on \href{https://gitlab.inf.ethz.ch/prj-cgl/cgl-ai-character/talking-to-your-data.git}{our GitLab repository}.}.

To ground the study in real-world behavior, we instantiated two distinct personas the participants are likely to resonate with, namely an office professional (Persona A; sedentary and sleep-deprived) and an active student (Persona B; physically active but stressed).
We then extracted longitudinal traces from the \emph{LifeSnaps} dataset~\cite{Yfantidou2022} to plausibly represent these personas.
To this end, we developed a custom scoring tool to extract representative 14-day segments based on persona-specific fitness functions as follows:

\begin{itemize} 
    \item \textbf{Persona A (Sedentary/Sleep-Deprived):} Windows were scored based on low average activity ($\mu\approx$5,000 steps), short sleep duration ($\mu\approx$5.5h), and a weekend "sleep rebound" effect. 
    \item \textbf{Persona B (Active/Stressed):} Windows were scored using a Gaussian fit for high activity levels ($\mu>$ 10,000 steps), high sleep variability, and elevated resting heart rate as a proxy for recovery stress. 
\end{itemize}

From the set of plausible traces, we selected two at random and enriched them with a narrative vignette describing the persona's lifestyle and habits as well as two health goals the participants were tasked to achieve during their interaction with our system (see Table~\ref{tab:personas}).
We defined a diagnostic and an actionable goal per persona to keep the conversation diverse~\ref{fn:repoRef}.
An example of the shared visual context during the user study is depicted in Figure~\ref{fig:interface}.

\begin{table}[t]
\renewcommand{\arraystretch}{1.3}
\centering
\caption{User Study Persona Profiles and Data Grounding\protect\footnotemark.}
\label{tab:personas}
\footnotesize
\begin{tabular}{@{}p{2.2cm}p{6.4cm}p{6.4cm}@{}}
\toprule
\textbf{Feature} & \textbf{Persona A: Office Professional} & \textbf{Persona B: Active Student} \\ \midrule
\textbf{Description} & Late 20s; full-time office job; sedentary; stays up late scrolling on phone. & 24-year-old student; active lifestyle; high-intensity workouts; exam stress. \\
\textbf{Data Trace} & $\mu \approx 5,518$ steps; $\mu \approx 5.75$h sleep ($5$h $44$m); weekend sleep rebound. & $\mu \approx 11,077$ steps; $\mu \approx 6.3$h sleep ($6$h $18$m); high variability. \\
\textbf{Diagnostic Goal} & Why do I feel tired during the day, even when I sleep in on weekends?  & Why do I feel sore or restless even though I exercise almost every day? \\
\textbf{Actionable Goal} & What are 1--2 specific changes could I try this week to improve both my activity level and sleep quality? & What should I adjust in my daily routine or training this week 	to recover better and reduce stress? \\ \bottomrule
\end{tabular}
\end{table}
\footnotetext{A detailed overview of the user study setup is available on \href{https://gitlab.inf.ethz.ch/prj-cgl/cgl-ai-character/talking-to-your-data.git}{our GitLab repository}.}

\subsection{Experimental Design and Procedure} \label{subsec:procedure}

We employed a 2x2 mixed factorial design. The \emph{Interface Condition} (Dashboard vs. Agent\footnote{Note that the Agent condition also includes the dashboard as explained in Section~\ref{subsec:system_overview}.
However, for the sake of readability, we will only refer to this condition as the \emph{Agent}.}) was treated as a within-subjects factor, while the \emph{Persona} (A vs. B) was counterbalanced across participants to prevent learning effects.
No participant saw the same persona or the same condition twice.

We recruited 5 participants (3 male, 2 female) from our institute's scientific staff.
Upon arrival, participants were briefed on the "Simulated Self" task and completed an entry questionnaire assessing familiarity with wearable sensors, statistical literacy, and conversational agents (all participants indicated high familiarity and literacy, except for P4 who indicated no prior use of wearable sensors).
The procedure consisted of two open-ended sessions ($\mu=16$ minutes per session):

\begin{enumerate}
    \item \textbf{Dashboard Condition:} Participants explored the static visualizations independently. To create a robust baseline for comparison, we employed a \textbf{concurrent think-aloud protocol}. This forced participants to externalize their reasoning, ensuring that any performance gains in the Agent condition were not merely due to the lack of verbal processing in the baseline.
    \item \textbf{Agent Condition:} Participants viewed the same charts while engaging in a "Shared-View" dialogue with a Unity-based embodied agent (called "Vincent"). Interaction was facilitated via a push-to-talk microphone. The agent utilized the dual-agent architecture described in Section~\ref{subsec:system_overview}, grounding its responses in a precomputed \emph{Insight JSON} to ensure objective consistency with the dashboard data.
\end{enumerate}

Between sessions, participants completed post-condition questionnaires (see Section~\ref{subsec:measures}).
The study concluded with a semi-structured exit interview comparing the two conditions in terms of understanding, actionability, mental demand, and trust.
All interactions were recorded and transcribed for qualitative coding.

\subsection{Measures and Analysis} \label{subsec:measures}

We collected three categories of data to evaluate the sensemaking process, namely subjective experience, objective actionability, and thematic analysis:

\begin{itemize}
    \item \textbf{Subjective Experience:} We utilized 7-point Likert scales to measure Perceived Understanding, Actionability, Trust, and Social Presence. These were administered via digital forms on the MacBook station.

    \item \textbf{Objective Actionability (Specificity Score):} To measure the depth of data reflection, we analyzed the health interventions proposed by participants at the end of each session. Each suggestion was rated on a 3-point author-defined Specificity Score used for qualitative comparison: 
    \begin{itemize}
        \item \textbf{0 (Vague):} Non-specific advice (e.g., "I should sleep more").
        \item \textbf{1 (Moderate):} General strategy (e.g., "I need to exercise on workdays").
        \item \textbf{2 (Concrete):} Data-driven, feasible plan (e.g., "Reduce screen time on Thursdays to offset the 2:00 PM energy dip identified in the sleep charts").
    \end{itemize}

    \item \textbf{Thematic Analysis:} Transcripts from both the think-aloud (Dashboard) and conversational (Agent) sessions were analyzed using thematic coding. This analysis focused on identifying differences in how participants verbalized statistical variance, correlations, and the "Persona Effect" across the two modalities.
\end{itemize}

Given the pilot nature of this study ($N=5$), the following analysis focuses on descriptive medians and the qualitative nuances of the participants' problem-solving trajectories leading up to their final health recommendations.
\section{Results and Discussion} \label{sec:results}

Given the pilot scale of the study, all quantitative results are reported descriptively and are intended to illustrate interaction patterns rather than support inferential claims.
Our analysis reveals that while the Dashboard facilitates detailed data scrutiny, the Agent condition significantly streamlines the transition from observation to concrete health planning by acting as a narrative bridge.

\subsection{Subjective Experience: Effort and Actionability} \label{subsec:subjective}

The post-condition Likert scales (see Figure~\ref{fig:likert}) reveal a significant shift in how participants perceived their interaction with the data.
The most striking result was the reduction in cognitive load; participants rated the Dashboard as requiring significantly higher mental concentration (Median 5.0) compared to the Agent (Median 2.0).

While perceived understanding remained high across both conditions (Median 6.0), the Agent condition yielded higher scores for the realism and specificity of identified actions (Median 7.0 vs. 6.0 for Dashboard).
However, Participant 2 warned that this offloading could induce a "passive" thought process, where the user might stop "learning to interpret signals from [their] own body" (Participant 2).
Trust in the agent's insights remained high (Median 7.0), validating the dual-agent design pattern as an effective method for providing objective, grounded reflection.

\subsection{Objective Actionability: From Themes to Tactics} \label{subsec:actionability}

The depth of reflection was quantified using a 3-point Specificity Score applied to the health plans proposed at the end of each session.

\begin{itemize} \item \textbf{Dashboard Specificity ($\mu$=1.25):} Independent inspection often led to broad "thematic" or reactive strategies. For example, Participant 3 proposed a general goal to "reduce the exercise a little bit", and Participant 4 focused on broad "batch cooking". \item \textbf{Agent Specificity ($\mu$=2.0):} Collaborative dialogue catalyzed "tactical" plans. Participant 1 translated sleep data into a concrete commitment to "read a book [...] for at least 30 minutes" to mitigate screen-induced sleep loss, while Participant 3 adopted specific "desk-friendly exercises like seated leg lifts" (Participant 3). \end{itemize}

This shift suggests that "spoken statistics" grounded in a shared visual context allow the agent to act as a co-interpreter, helping users move past outliers to reach concrete conclusions.

\subsection{The Future Use Paradox and Embodiment} \label{subsec:comparative}

\begin{figure}[t] 
    \centering 
    \begin{subfigure}[b]{0.49\textwidth} 
        \centering \includegraphics[width=\textwidth]{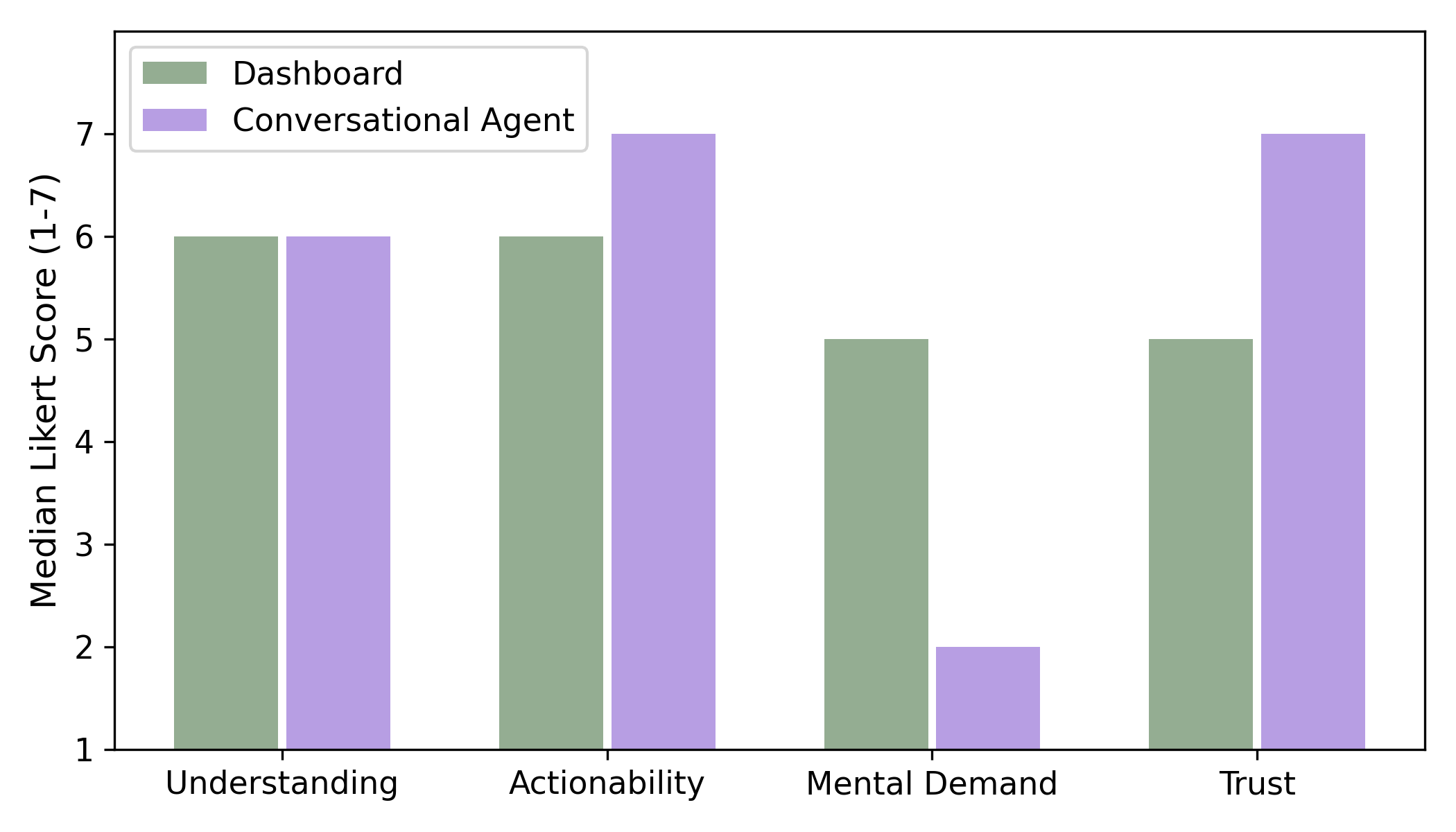} 
        \caption{Median Likert scores from post-condition questionnaires.} 
        \label{fig:likert} 
    \end{subfigure} 
    \hfill 
    \begin{subfigure}[b]{0.49\textwidth} 
        \centering \includegraphics[width=\textwidth]{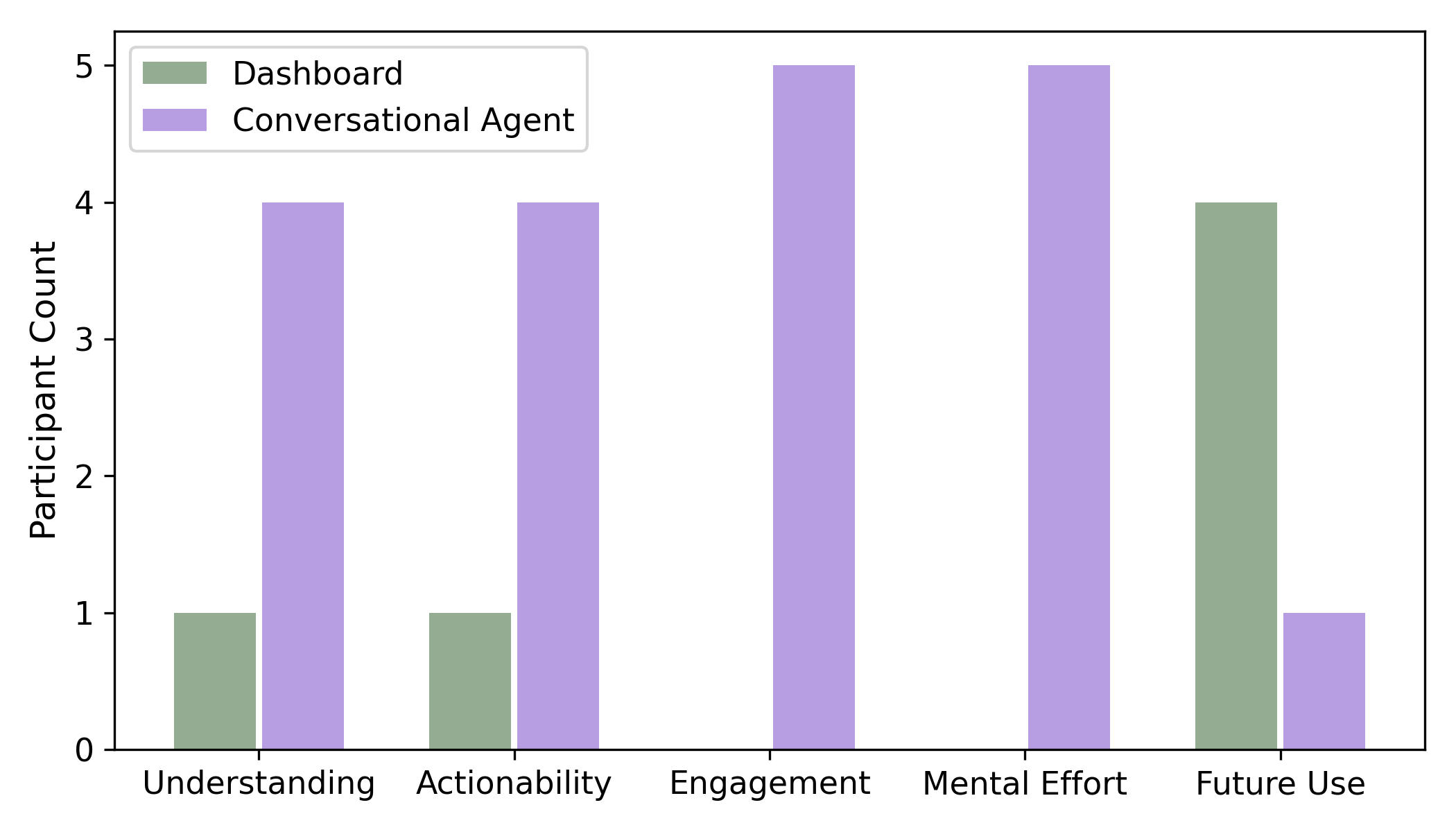} \caption{Comparative preferences from the exit questionnaire.} 
        \label{fig:paradox} 
    \end{subfigure} 
    \caption{Quantitative evaluation of Dashboard vs. Conversational Agent. While the Agent reduces effort and increases actionability, users default to the Dashboard for initial data exploration.} \label{fig:quantitative_results}
\end{figure}

Despite the Agent's superior performance in actionability, the exit questionnaire (Figure~\ref{fig:paradox}) revealed a "Future Use Paradox."
Most participants (4 out of 5) indicated they would still prefer a Dashboard for their first interaction with their own wearable data.
This indicates that users value the analytical control of a dashboard for initial exploration, viewing the Agent as a secondary tool for deeper reflection.

This preference may be linked to the "split-attention" effect observed with the 3D avatar.
While some felt the avatar helped "bridge the gap" (Participants 2, 3), others found it distracting.
Participant 1 noted she felt "forced to look at the agent" rather than "studying the data at the same time" (Participant 1).
Consequently, the effectiveness of the Agent was driven not by visual presence, but by verbal deictic cues (e.g., "last Wednesday") which successfully established joint attention without the need for physical pointing.

\subsection{Cognitive Trajectories in Sensemaking}

Thematic analysis of transcripts revealed a shift in how participants processed data.
In the Dashboard condition, users exhibited an \emph{extrema effect}, where reasoning was anchored to outliers.
Participant 1 noted, "I was mostly looking at extrema," leading to reactive conclusions. 

In contrast, the Agent facilitated \emph{trend synthesis}.
By providing "spoken statistics" grounded in temporal anchors (e.g., "last Wednesday"), the agent reduced the cognitive load required to identify correlations.
This shifted the user from a manual investigator to an active participant in a reflective dialogue.
However, as described in Section~\ref{subsec:comparative}, a "Future Use Paradox" emerged where 4 out of 5 participants still preferred the Dashboard for primary data exploration, viewing the Agent as a secondary tool for deep reflection.

\subsection{The Role of Embodiment and Verbal Grounding}

Contrary to our hypothesis, the 3D avatar's physical presence was polarizing.
Participant 1 felt "forced to look at the agent" when she preferred studying the data, suggesting that embodied avatars can create a cognitive distraction by competing for visual attention.
Participants 2, 4, and 5 mentioned that they did not pay much attention to the visual representation of the avatar but found its voice helpful while looking at the data simultaneously.
Contrarily, Participant 3 found the embodiment to "add value to the interaction."

Crucially, the system's effectiveness was driven by \emph{verbal deictic cues}.
Using phrases like "weekend rebound" established joint attention without requiring physical pointing.
This indicates that for Health-IUI, precise verbal grounding might be more critical for sensemaking than visual embodiment. 
These findings suggest that the contribution of embodiment in this prototype lies primarily in supporting shared attention and conversational grounding around the data, rather than in the visual form of the avatar alone.

\subsection{Design Implications and Limitations}

Based on these findings, we propose that the separation of the Observer (data analysis) and Presenter (dialogue) remains essential also for health interface safety.
Grounding the interaction in a precomputed \textit{Insight JSON} ensures that social presence is balanced with factual integrity, effectively reducing the risk of AI hallucinations.
Furthermore, designers must account for a "Data-Literacy Threshold."
For users with high data-literacy, such as Participant 3, proactive agent interpretation felt less empowering than independent exploration. 
We therefore frame all design implications as preliminary hypotheses to guide future, larger-scale investigations rather than definitive conclusions.

As an exploratory pilot ($N=5$), several limitations remain.
While the Simulated Self protocol maintained experimental control, it lacked the emotional stakes of personal health data.
Future studies should evaluate how this paradigm scales when navigating months of noisy data to determine if the increased specificity in action planning is sustained over time.
Furthermore, our study utilized a cross-sectional design that captured only a single moment of reflection.
Future work should evaluate how this paradigm scales over longitudinal contexts; specifically, whether the increased specificity in action planning is sustained over months of noisy, real-world data or if a "novelty effect" leads to decreased engagement with the agent over time.
\section{Conclusion}
\label{sec:conclusion}

We presented a system for "talking to your data" through an embodied agent that facilitates objective reflection.
By separating data analysis from conversational delivery via a dual-agent architecture, we demonstrated a path toward intelligent health interfaces that are both engaging and factually grounded.
These exploratory insights suggest that conversational reflection may help lower the interpretation gap of traditional dashboards, turning personal data into a meaningful dialogue.

\begin{acknowledgments}
    This research was funded by the \emph{Hasler Foundation}\footnote{https://haslerstiftung.ch} under Research Grant No. 2025-01-13-250.
\end{acknowledgments}

\section*{Declaration on Generative AI}

During the preparation of this work, the author(s) used Writefull and ChatGPT in order to: Grammar and spelling check. After using these tool(s)/service(s), the author(s) reviewed and edited the content as needed and take(s) full responsibility for the publication’s content.



\end{document}